# Field-effect transistors and intrinsic mobility in ultra-thin MoSe$_2$ layers

S. Larentis, B. Fallahazad, and E. Tutuc

*Microelectronics Research Center, Department of Electrical and Computer Engineering,*
*The University of Texas at Austin, Austin, TX78758, USA*

We report the fabrication of back-gated field-effect transistors (FETs) using ultra-thin, mechanically exfoliated MoSe$_2$ flakes. The MoSe$_2$ FETs are *n*-type and possess a high gate modulation, with On/Off ratios larger than $10^6$. The devices show asymmetric characteristics upon swapping the source and drain, a finding explained by the presence of Schottky barriers at the metal contact/MoSe$_2$ interface. Using four-point, back-gated devices we measure the intrinsic conductivity and mobility of MoSe$_2$ as a function of gate bias, and temperature. Samples with a room temperature mobility of ~50 cm$^2$/V·s show a strong temperature dependence, suggesting phonons are a dominant scattering mechanism.

Transition metal dichalcogenides (TMDs) are materials characterized by a MX$_2$ formula where M stands for a transition metal (Mo, W), and X stands for a chalcogen (S, Se or Te)[1,2]. The transition metal dichalcogenides are layered materials, consisting of layers with an X-M-X structure. Within each layer the M-X bonds are covalent, while separate layers are bonded via Van der Waals interaction. Recent studies[3–6] proved that micromechanical exfoliation, employed to isolate graphene monolayers[7] is also an effective method to obtain thin flakes of various TMDs. The layers can be indentified[4–6] using optical microscope thanks to the thickness dependent contrast on SiO$_2$/Si substrates. Various TMD-based electronic devices have been demonstrated, with MoS$_2$ being investigated most extensively to date. Examples include MoS$_2$ monolayer *n*-type field effect transistors (FETs)[8,9], MoS$_2$ photo-transistor[10], MoS$_2$ bilayers chemical sensors[11], logic gates, memory cells and amplifiers using large MoS$_2$ flakes[12–14]. In addition, top-gated *p*-type FETs using monolayer WSe$_2$[15], and ambipolar back-gated FETs using multilayered WS$_2$[16] have been demonstrated. The TMD-based FETs generally possess a high On/Off ratio, larger than $10^6$. Such semiconductors can be scaled down to a 0.7 nm thickness monolayer, which renders them attractive as an ultra-thin body for aggressively scaled FETs. The larger band-gap by comparison to graphene can ensure a large On/Off ratio.

In this study we examine the electron transport in another TMD, namely MoSe$_2$, a semiconductor with an indirect band-gap of 1.1 eV[17,18] in bulk, which increase to 1.55 eV and become direct in monolayer and bilayer[18,19]. We discuss the realization of field-effect transistors using ultrathin MoSe$_2$ flakes mechanically exfoliated on SiO$_2$/Si substrates. The MoSe$_2$ FETs are *n*-type, and possess a good gate control, with On/Off ratios larger than $10^6$. Four-point, gated devices allow us to measure the intrinsic conductivity, and extract the mobility of MoSe$_2$. The temperature (*T*) dependence reveals that the mobility increases significantly with decreasing the temperature, suggesting that phonon scattering dominates at room temperature.

The MoSe$_2$ flakes used here are produced by micromechanical exfoliation using commercially available MoSe$_2$ powder (Materion Inc.) with grain size < 44 µm (mesh -325). The MoSe$_2$ flakes are exfoliated on a 285 nm-thick SiO$_2$ film, thermally grown on highly doped *n*-type Si (100) wafers (N$_D$ > $10^{20}$ cm$^{-3}$). The exfoliated flakes are identified using optical microscope, and their thickness measured by atomic force microscopy (AFM). Figure 1(a) shows an optical micrograph of a MoSe$_2$ flake on SiO$_2$/Si substrate, while Fig. 1(b) shows the flake topography, probed by AFM, illustrating layered staking. The typical flake size is 1-3 µm, and their thickness ranges between 3 to 80 nm. Individual MoSe$_2$ flakes typically exhibit terraces with various thicknesses, with a terrace surface roughness ranging from 0.2 to 0.6 nm. The optical contrast of the MoSe$_2$ on 285 nm thick SiO$_2$ substrate allows identification of flakes as thin as 4.8 nm (7 layers)[4–6].

To assess the MoSe$_2$ quality we performed X-ray diffraction (XRD) on powder samples. The power XRD pattern, shown in Fig. 1(c) matches with the 2H-Drysadallite patterns[20] ensuring our material is characterized by hexagonal 2H-MoSe$_2$ structure with space group D$^4_{6h}$ (P6$_3$/mmc)[21], and in agreement with previous MoSe$_2$ powder XRDs[22,23] studies. The full width at half maximum (FWHM) of the <002> peak is 0.223°. To further assess the exfoliated MoSe$_2$ flakes we performed µ-Raman spectroscopy using a Renishaw InVia Raman microscope. Figure 1(d-e) show Raman spectra using 442 nm [Fig. 1(d)] and 532 nm [Fig. 1(e)] excitation wavelengths on a MoSe$_2$ flake. The spectrum of Fig. 1(d), acquired with an incident laser power of 30 mW and 650 nm spot size presents four peaks at 169, 242, 285 and 352 cm$^{-1}$, corresponding to the E$_{1g}$, A$_{1g}$, E$^1_{2g}$ and A$^2_{2u}$ modes respectively, as reported by Sekine *et al.*[24] and more recently by Tongay *et al.*[18]. We note that the A$^2_{2u}$ mode (352 cm$^{-1}$) is an infrared active mode, not present in Raman scattering in bulk samples. The emergence of this mode in Raman scattering acquired on small flakes suggests a breakdown of inversion symmetry, possibly because of the substrate. A similar observation has been made recently in a Raman spectroscopy study of Bi$_2$Te$_3$ nanoplates[25]. The Raman spectrum of Fig. 1(e), acquired with a 5 mW incident laser power shows similar peaks as Fig. 1(d) data, but with a higher intensity of the A$_{1g}$ peak (242 cm$^{-1}$) with respect to the other modes. This observation is consistent with previous studies which show that the A$_{1g}$ mode is better resolved at longer excitation wavelengths.[24]

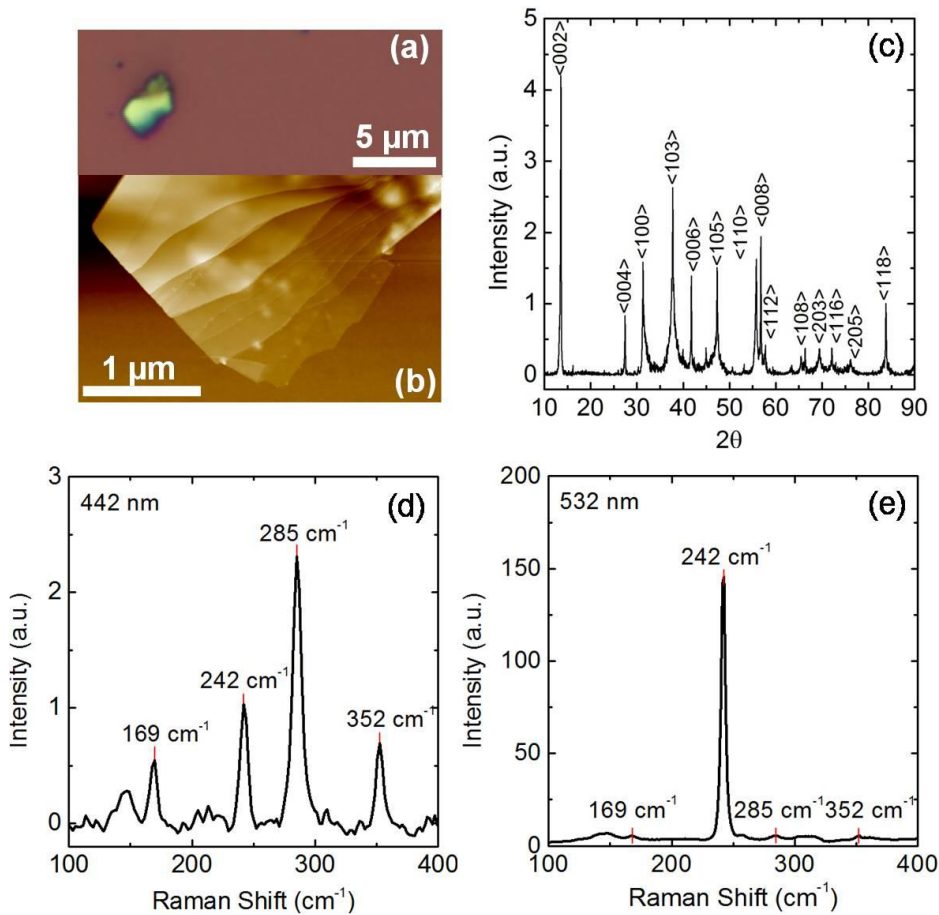

Fig. 1. (color online) (a) Optical micrograph of a MoSe$_2$ flake on SiO$_2$/Si substrate. (b) AFM topography of the thinner section (darker section in the optical image) of the same flake shown in panel (a). The data illustrates layered staking in MoSe$_2$ flakes. (c) MoSe$_2$ powder XRD data. (d-e) Raman spectra acquired on a MoSe$_2$ flake using excitation wavelengths of 442 nm [panel (d)] and 532 nm [panel (e)].

The FWHM of E$_{1g}$, A$_{1g}$, E$^1_{2g}$ and A$^2_{2u}$ Raman peaks of Fig. 1(d) are 5.5, 5.2, 7, and 5 cm$^{-1}$ respectively. The A$_{1g}$ peak FWHM of Fig. 1(e) is 3.5 cm$^{-1}$, assuming in all cases a Lorentzian fit. The higher FWHM values when using a 442 nm excitation wavelength is explained by the higher incident laser power. Previous Raman spectroscopy studies on MoS$_2$ shows the same trend, explained by enhanced thermal effects[5,26].

In order to electrically probe an exfoliated MoSe$_2$ flake, we select a single terrace with uniform thickness to define the active region of the field-effect transistor. Electron-beam lithography (EBL) combined with reactive ion etching using Cl$_2$ are then used to define the device active region. The metal contacts are defined by a second EBL step followed by 80 nm-thick Ni evaporation and lift off. The highly doped Si substrate serves as back-gate for the MoSe$_2$ FETs. Several two- and four-point back-gated field effect transistors (FET) were investigated in this study.

The output and transfer characteristics of an FET fabricated on a 5.8 nm thick MoSe$_2$ flake (~ 8 layers), with channel lenght $L$= 1.8 μm and width $W$ = 0.8 μm are shown in Fig. 2. The measurements presented in this paper are carried out in vacuum (~10$^{-7}$ torr), and in dark. In each measurement the source contact is grounded, while the drain is biased. Figure 2(a) shows transfer characteristics defined as drain current ($I_D$) vs. drain voltage ($V_D$) measured at various back gate voltages ($V_G$) at room temperature. The $I_D$-$V_D$ dependence is mostly linear, and does not show saturation at high drain bias, in contrast to what is expected in a conventional FET. Moreover the $I_D$-$V_D$ data exhibit a slight super-linear behavior at low drain bias, which suggests that the electrons are injected through a Schottky barrier at metal (Ni)-semiconductor interface. Figure 2(b) shows the transfer characteristic ($I_D$-$V_G$) measured at two different drain biases $V_D$ = 50 mV and $V_D$ = 1 V. The device shows an $n$-type behavior, and is depleted from free carriers for $V_G$ < 0 V. The threshold voltage of this device, $V_T \cong 0$ V does not appear to change noticeably with $V_D$, consistent with a long-channel device. The $I_{on}/I_{off}$ ratio at $V_D$ = 1 V curve is larger than 10$^6$, similar to reported On/Off ratio values for MoS$_2$, WS$_2$, and WSe$_2$ devices[8,15,16], and explained by the large energy gaps that characterize this family of materials.

To further probe the electron injection in MoSe$_2$, in Fig. 3 we show two sets of output characteristics measured on the same two-point back-gated MoSe$_2$ FET, but using a different source contact in each data set. For the same $V_G$ value, different $I_D$ values are obtained depending on which physical contact is used as source. This asymmetry in $I_D$–$V_D$ data further confirms the presence of a Schottky barrier at the metal-semiconductor contact. As a result the electron injection depends not only to the device geometry, e.g. contact area, but also on the electric field across the metal/MoSe$_2$ interface, and therefore will be sensitive to MoSe$_2$ flake thickness, SiO$_2$ dielectric thickness, as well as gate and drain bias. Schottky barrier contacts are common place for other nano-electronic devices, such as carbon nanotube (CNT)[27] and nanowire devices[28]. The presence of non-ohmic contacts affects adversely the device performance by reducing the On state current and prevents a quantitative analysis of the device characteristics. Most importantly, extracting the intrinsic MoSe$_2$ mobility from data such as that of Fig. 2 is difficult.

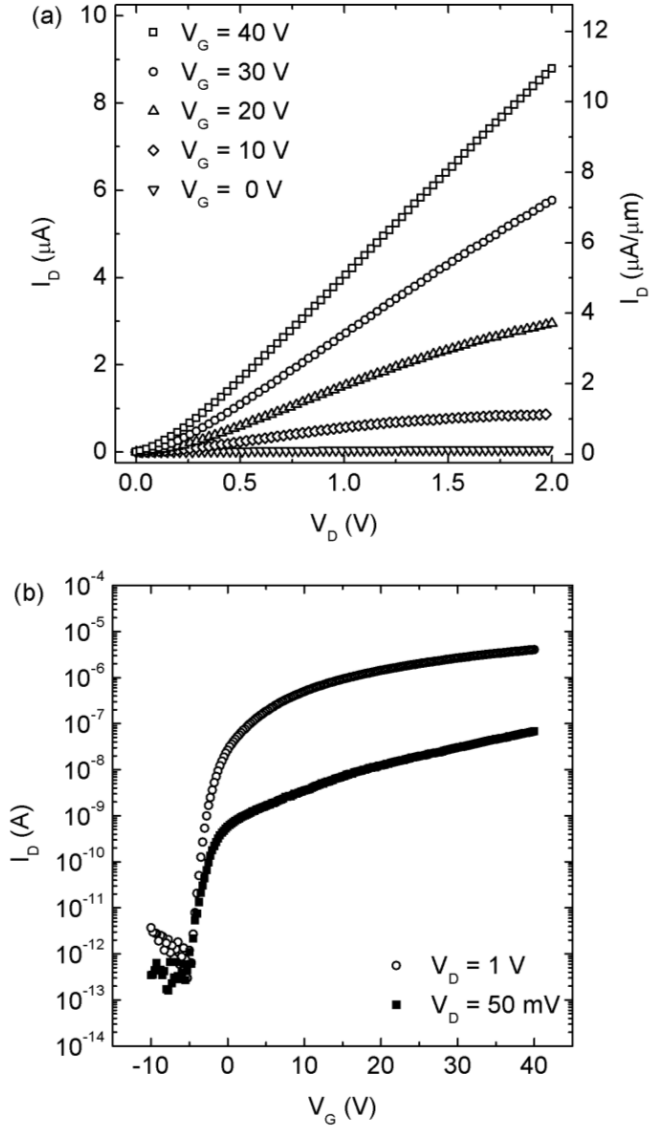

Fig. 2. (a) $I_D$ vs. $V_D$ measured at different $V_G$ values. The data show a super-linear behavior at low $V_D$ suggesting the presence of a Schottky barrier at the metal MoSe$_2$ contact. (b) $I_D$ vs. $V_G$ traces measured at $V_D = 50$ mV (solid squares), and $V_D = 1$ V (open circles) with $I_{on}/I_{off} > 10^6$ at $V_D = 1$ V.

To probe the intrinsic mobility of MoSe$_2$ flakes, we fabricate four-point back-gated devices, which allow conductivity measurements without contributions from the contact resistance of the metal-semiconductor Schottky barriers. The inset of Fig. 4 (a) shows an AFM image of a four-point MoSe$_2$ device. The outer contacts labeled *S* and *D* serve as source and drain, respectively. The inner contacts ($V_1$, $V_2$) are used as voltage probes, and have a limited overlap with the MoSe$_2$ flake to minimize screening of the gate-induced charge density in the channel. The measured channel conductance ($G$) is defined as $G = I_D/(V_1-V_2)$. Figure 4(a) shows the $G$ vs. $V_G$ data measured at different $T$ values from 298 to 78 K. For $V_G$ values lower than a threshold voltage ($V_T$), $G$ remains vanishing. Above threshold, the $G$ increases with $V_G$, with an approximately linear dependence. As the temperature is reduced, $V_T$ shifts progressively towards higher voltages.

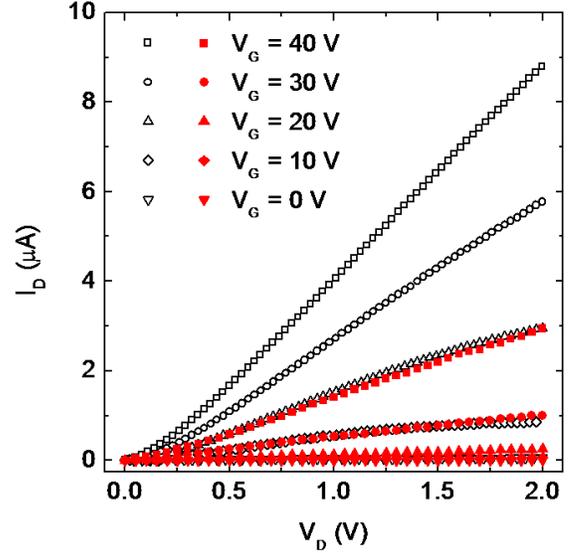

Fig. 3. (color online) Different $I_D$-$V_D$ data sets obtained by swapping drain and source contacts on the same device. The asymmetry of the $I_D$ values when swapping the S and D contacts is characteristic of FETs with Schottky contacts.

To offset the $V_T$ shift with $T$, Fig. 4(b) shows $G$ vs. $V_G$-$V_T$, at different $T$ values. Figure 4(b) data show a noticeable increase of the $dG/d(V_G-V_T)$ slope with decreasing $T$. The intrinsic mobility can be thus extracted as $\mu = (L/W) \cdot dG/dV_G \cdot C_{ox}^{-1}$, where $C_{ox} = 1.2 \times 10^{-8}$ F·cm$^{-2}$ is the capacitance of the 285 nm-thick bottom SiO$_2$ dielectric; $L$ and $W$ denote the length and width of the active device area, respectively. The inset of Fig. 4 (b) shows the mobility dependence on temperature for three different devices. The room temperature mobility is as high as 50 cm$^2$/V·s, and increases almost four fold when reducing the temperature to 78 K. In two dimensions, charged impurity scattering causes a temperature independent mobility in the degenerate limit, as in the case of graphene[29], and a $\mu \propto T$ dependence for a non-degenerate two-dimensional electron system. Acoustic phonon scattering generates a $\mu \propto T^{-1}$ dependence, and optical phonon scattering, including polar optical phonons, cause a stronger $T$-dependence in TMDs[30]. A functional fit of the form $\mu^{-1}(T) = A + BT^\nu$ to the $\mu$ vs. $T$ data of Fig. 4(b) inset yields an exponent $\nu=2.1$ for the highest mobility sample, suggesting that phonon scattering dominates in this sample. Temperature dependent Hall mobility measurements carried out on bulk MoSe$_2$ samples also report an increased mobility with reducing temperature.[31]

Lastly we address the contact resistance in our devices. Having measured the MoSe$_2$ resistivity using four-point devices, we can subtract the flake intrinsic resistance from the measured source-to-drain resistance in order to estimate the contact resistance. The room temperature contact resistance values range between 200 kΩ at large, $V_G$-$V_T = 35$ V gate overdrive, and increase to 6 MΩ as $V_G$ approaches $V_T$. Reducing the temperature leads to an increase in contact resistance. The strong $V_G$ dependence of the contact resistance provides further evidence for the presence of a Schottky barrier at the metal/MoSe$_2$ interface, an obstacle which will have to be overcome in order to increase the On current.

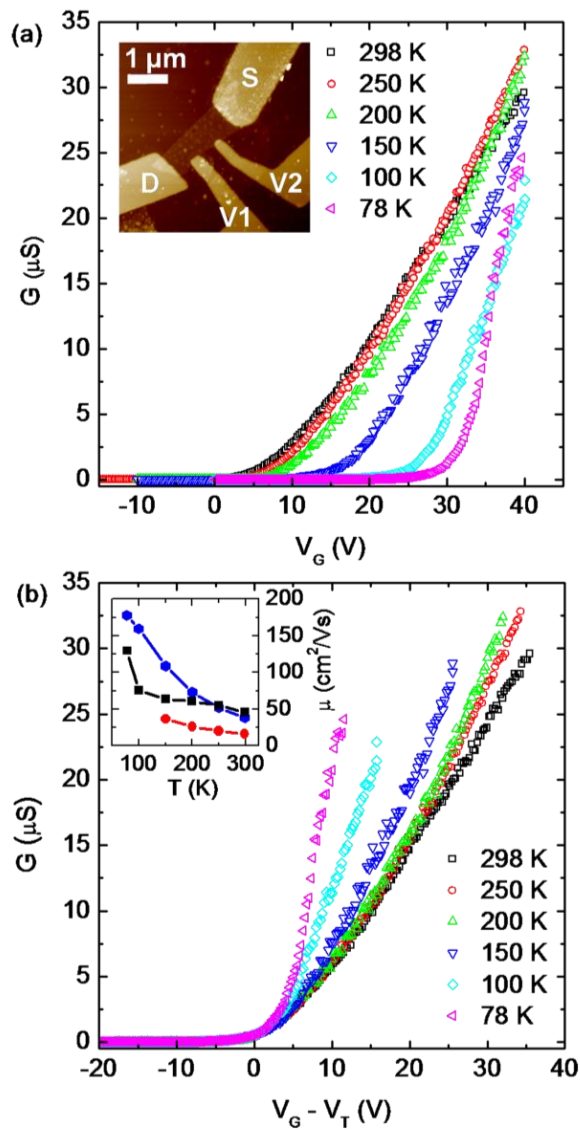

Fig. 4.(color online) (a) $G$ vs. $V_G$ at different temperatures, ranging from 298 to 78 K. The data shows a linear dependence above threshold, combined with a $V_T$ shift towards higher voltages when reducing $T$. Inset: AFM image of a four-point device. The source ($S$), drain ($D$), and voltage probes ($V_1$, $V_2$) are marked accordingly. (b) $G$ vs. $V_G$-$V_T$ data for the same set of temperatures. Inset: $\mu$ vs. $T$ for three different MoSe$_2$ devices.

In summary, we demonstrate $n$-type field-effect transistors on ultra-thin MoSe$_2$ flakes, showing high $I_{on}/I_{off}$ ratios. We probe the intrinsic mobility as a function of temperature using gated four-point device structures, and show that the mobility increases significantly with lowering the temperature, which suggest phonon scattering plays a dominant role at room temperature.

This work is supported by ONR and Intel. We thank R. Pillarisetty for discussions.